\documentclass{PoS}

\usepackage{comment}
\usepackage{amsmath}
\usepackage{amsfonts}
\usepackage{amssymb}%

\newcommand{\be}{\begin{equation}}
\newcommand{\ee}{\end{equation}}
\newcommand{\bdm}{\begin{displaymath}}
\newcommand{\edm}{\end{displaymath}}
\newcommand{\btb}{\begin{tabular}}
\newcommand{\etb}{\end{tabular}}
\newcommand{\bfig}{\begin{figure}}
\newcommand{\efig}{\end{figure}}

\renewcommand{\d}{\partial}
\newcommand{\ds}{\mathrm{d}}

\newcommand{\dd}{D\bar{D}}

\title{Line-shape analysis of charmonium resonances}

\ShortTitle{Line-shape analysis}

\author{\speaker{Susana Coito}\\
        Institute of Physics, Jan Kochanowski University, 25-406 Kielce, Poland\\
        E-mail: \email{scoito@ujk.edu.pl}}

\abstract{We discuss weather the new enhancements found by BES, alias the $Y(4220)$, $Y(4260)$, $Y(4360)$, and $Y(4390)$ are true resonances. We argue that the nearby thresholds $D_s^*\bar{D}_s^*$, $D\bar{D}_1+\bar{D}D_1$, $D_s\bar{D}_{s1}+\bar{D_s}D_{s1}$ and $D^*\bar{D}_1+\bar{D}^*D_1$, as well as the $\psi(4160)$ and $\psi(4415)$ states have a strong influence over the observed $ J/\psi \pi^+\pi^-$ and $h_c \pi^+\pi^-$ line-shapes. We propose an unitarized effective Lagrangian model to study the dynamical effect of the interaction between each known $\psi$ state and its closest thresholds. In addition, we present some of our recent motivating results, using the same model, for the $\psi(3770)$ resonance, where the distortion from a Breit-Wigner line-shape is shown to result not only from the kinematic interference, but also from the influence of the $D^0\bar{D}^0+D^+D^-$ one-loops. Moreover, two poles were found, at about 3.78 GeV and at 3.74 GeV, the second one generated dynamically, yet contributing to the distortion of the line-shape.}

\FullConference{XVII International Conference on Hadron Spectroscopy and Structure - Hadron2017\\
		25-29 September, 2017\\
		University of Salamanca, Salamanca, Spain}

\begin{document}


\section{Introduction}

Recently, the BES collaboration announced the detection of four new vectorial charmonium states: two in the process $e^+e^-\to J/\psi \pi^+\pi^-$ \cite{prl118p092001}, at
\begin{align}
\label{jpsi}
&Y(4260):\ M=4222.0\pm 3.1\pm 1.4,\ \Gamma=44.1\pm 4.3\pm 2.0\ \mathrm{MeV}\ ,\\
&Y(4360):\ M=4322.0\pm 10.4\pm 7.0,\ \Gamma=101.4^{+25.3}_{-19.7}\pm 10.2\ \mathrm{MeV}\ ,
\label{jpsib}
\end{align}
and two in the process $e^+e^-\to h_c \pi^+\pi^-$ \cite{prl118p092002}, at
\begin{align}
&Y(4220):\ M=4218.4^{+5.5}_{-4.5}\pm 0.9,\ \Gamma=66.0^{+12.3}_{-8.3}\pm 0.4\ \mathrm{MeV}\ , \\
&Y(4390):\ M=4391.5^{+6.3}_{-6.8}\pm 1.0,\ \Gamma=139.5^{+16.2}_{-20.6}\pm 0.6\ \mathrm{MeV}\ .
\label{hc}
\end{align}
The $Y(4260)$ enhancement was identified with the Particle Data Group (PDG) entry $X(4260)$ \cite{pdg}, and the $Y(4360)$ with the entry $X(4360)$, while the other two are claimed to be new resonances. In this work, we point out that such conclusions are too rash and may be incorrect, since they are drawn from mere ``bump hunting'' using naive Breit-Wigner fits, without taking into account the nonresonant production due to nearby Okubo-Zweig-Iizuka(OZI)-allowed thresholds. Concerning the $X(4260)$ enhancement, it has been pointed out in severeal works, see Refs.~\cite{prd79p111501,prl105p102001}, that it is not a true state, but a mere overall bump generated by a more detailed and complex structure. In fact, the very $X(4260)$ has not been seen in any of the OZI-allowed decay channels, in particular the $D_s^*\bar{D}_s^*$, that falls right on the resonance position \cite{pdg}. Also, different decay channels should have different line-shapes, as the continuum is different, so one should look for correlations among the structures seen in different channels, rather than conclude them to be independent resonances. 

In order to clarify the new peaks seen in Refs.~\cite{prl118p092001} and \cite{prl118p092002}, we present an effective Lagrangian model to check whether the observed enhancements can be originated by interferences among the already known resonance $\psi(4160)$ and thresholds $D_s^*D_s^*$ and $DD_1$ located above, and resonance $\psi(4415)$ and thresholds  $D_sD_{s1}$ and $D^*D_1$, also above (note that here and henceforth, we omit the $bar$ for antiparticles to simplify notation). Although the very $\psi(4160)$ and $\psi(4415)$ states are nearly closed in such channels, their ``tails'', which result not only from kinematic interferences, but possibly from dynamical effects as well, could still be seen in such channels, and with higher cross sections than those measured in channels $J/\psi \pi^+\pi^-$ and $h_c \pi^+\pi^-$. Such measurements would help to disentangle the nature of the peaks, and to clear out some theoretical speculation over states whose existence has not been properly examined.

The idea that a nearby resonance can leave a ``tail'' on closed decay channels has been worked out before in different charmonium systems, such as the prediction of a scalar around 3.7 GeV which can be seen in channel $DD$, and of an axial-vector in channel $D\bar D^*$ \cite{epja36p189}. Here, we present a method to study the mentioned states within an unitarized Lagrangian approach, similar to the one employed in Ref.~\cite{1708.02041,paper}. Yet, since the present study is preliminary, calculations are intended to be carried out elsewhere \cite{fpaper}. In addition, we show our recent result on the line-shape of the vector charmonium $\psi(3770)$, where, including, the pole structure was examined. Our optimistic results encourages us to carry on with the present research, on such a relevant issue to mesonic spectroscopy, as the nature of the $XYZ$ states.  


\section{Bumps and Thresholds in the experiment}

The PDG lists six vector charmonium states, namely the $J/\psi$, $\psi(2S)$, $\psi(3770)$, $\psi(4040)$, $\psi(4160)$, and $\psi(4415)$, respectively the $1S$, $2S$, $1D$, $3S$, $2D$, and $4S$ radial states. While there is still some debate concerning the quantum numbers of the $\psi(4415)$, three additional vectors are listed in PDG, namely the $Y(4230)$, $Y(4260)$, and $Y(4360)$ (called $X$ in the PDG), concerning which the isospin and $G$-parity have not been determined. It is tempting, therefore, to identify some of the claimed new resonances in Eqs.~\eqref{jpsi}-\eqref{hc} with such states. However, care is needed before drawing such conclusions, before a taking into account the influence of the dominant OZI-allowed decay channels, greatly responsible for the large widths of some resonances, e.g.~the $\rho(770)$ in its decay to $\pi\pi$ \cite{pdg}. Indeed, nonperturbative effects can contribute to the distortion of line-shapes, generation of new peaks, or misplaced bumps and dips in certain decay modes \cite{ppnp67p449}. As an example, the $\psi(4040)$ and $\psi(4160)$ appear as dips in the $D\bar{D}$ channel in data from Belle \cite{prd77p011103}, while in the same data, the $\psi(3770)$ and $\psi(4415)$ rise as bumps. In turn, the $X(4260)$ enhancement, seen at first in $J/\psi \pi^+\pi^-$ channel at BaBar \cite{prl95p142001}, rather than a true resonance, might result from the interference among several nearby thresholds, namely the $D_sD_s^*$, $D_s^*D_s^*$, and the resonances $\psi(4160)$ and $\psi(4415)$ \cite{prl105p102001}, with the peak falling right on the $D_s^*D_s^*$ position \cite{prd79p111501}.

\begin{figure}
\centering
\resizebox{!}{220pt}{\includegraphics{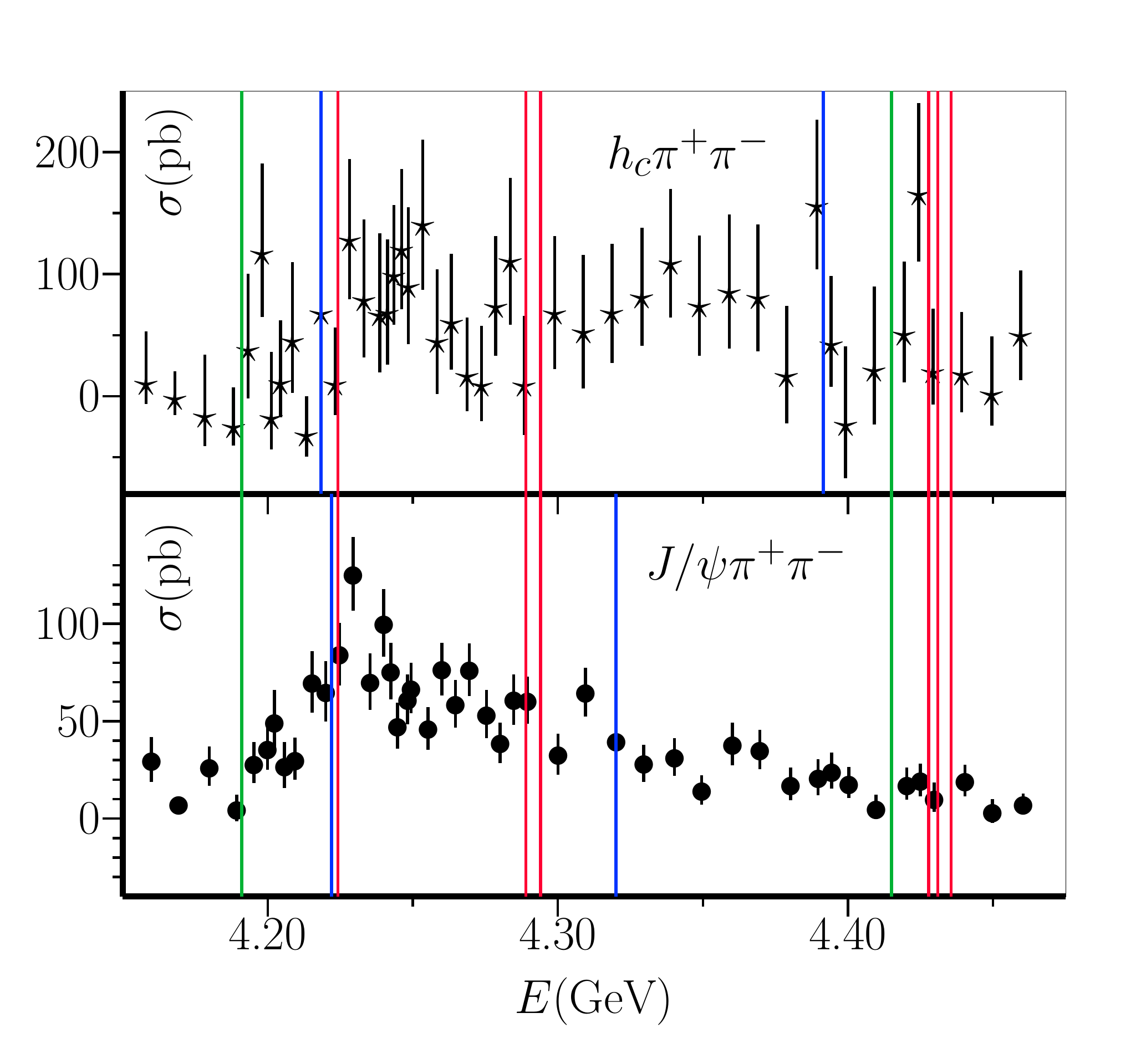}}
\caption{\label{fig1} Data from BES in Refs.~\cite{prl118p092002} (above) and \cite{prl118p092001} (below). Green lines: $\psi(4160)$ and $\psi(4415)$; Red lines: $D_s^*D_s^*$, $DD_1$, $DD_1'$,  $D_sD_{s1}$, $D^*D_1$, and $D^*D_1'$ thresholds (cf.~Table \ref{table}). Blue lines: $Y(4220)$, $Y(4260)$, $Y(4360)$, and $Y(4390)$ (cf.~Eqs.~\eqref{jpsi}-\eqref{hc}).}
\end{figure}

In Fig.~\ref{fig1} we plot the data in Refs.~\cite{prl118p092001} and \cite{prl118p092002}, together with the sharp position of the peaks in Eqs.~\eqref{jpsi}-\eqref{hc}, the position of the $\psi(4160)$ and $\psi(4415)$ resonances, and the position of the nearby OZI-allowed thresholds $D_s^*D_s^*$, $DD_1$, $DD_1'$,  $D_sD_{s1}$, $D^*D_1$, and $D^*D_1'$ (see Table \ref{table} for the exact masses). Intuitive conclusions may be drawn from the figure: i) although the overall line-shape in both channels is different, one can observe dips at the same energies, namely at the position of $\psi(4160)$ and $\psi(4415)$, and enhancements at the same threshold positions; ii) in some cases, bumps and dips are interchanged in both channels, such as at 4.27, 4.31 or 4.35 GeV, which might indicate competition between them; iii) also, there is no apparent reason why the enhancement at about 4.26 GeV in each channel should not be the same, since both channels represent decays of the vector charmonia; iv) it is not clear at all why there should be the resonances $Y(4360)$ and $Y(4390)$ in Eqs.~\eqref{jpsib} and \eqref{hc}, while discarding other structures in the spectra, specially in the $h_c\pi^+\pi^-$ distribution. The posed question is then, how theory can help to make sense of the puzzling experimental observations. We propose an effective Lagrangian model, as we describe below.

\begin{table}[b]
\centering
\begin{tabular}{c|cc}
Lines&Mass (MeV)&Width (MeV)\\
\hline
$\psi(4160)$&$4191\pm 5$&$70\pm 10$\\
$D_s^*D_s^*$&4224.2&$<3.8$\\
$DD_1$&4289&$31.7\pm2.5$\\
$DD_1'$&4294&$384^{+107}_{-75}\pm 74$\\
$\psi(4415)$&$4421\pm 4$&$62\pm 20$\\
$D_sD_{s1}$&4427.77&$<3.5$\\
$D^*D_1$&4431&$31.7\pm 2.5^{+2.1}$\\
$D^*D_1'$&4435.55&$384^{+109.1}_{-75}\pm 74$\\
\hline
\end{tabular}
\caption{\label{table}Resonance and threshold masses in PDG \cite{pdg}.}
\end{table}


\section{An Effective Model}

\begin{figure}
\centering
\begin{tabular}{c}
\resizebox{!}{20pt}{\includegraphics{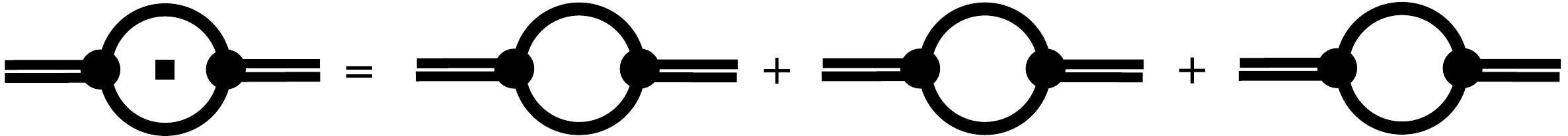}}\\[2mm]
\vspace*{2mm}
\resizebox{!}{20pt}{\includegraphics{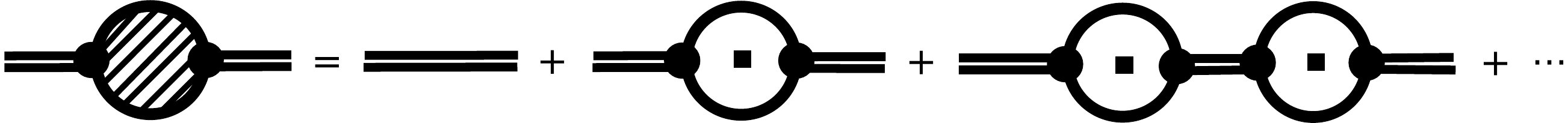}}
\end{tabular}
\caption{\label{fig2}One-loop, including three channels (above). Full $\psi$ propagator expansion (below).}
\end{figure}

As stated above, we aim to study weather the enhancements seen in the $J/\psi \pi^+\pi^-$ and $h_c \pi^+\pi^-$ cross section distributions may be threshold enhancements generated by interference among the $\psi(4160)$ resonance and thresholds $D_s^*D_s^*$, $DD_1$, and $DD_1'$, and among the $\psi(4415)$ and thresholds $D_sD_{s1}$, $D^*D_1$, and $D^*D_1'$. For the first process the vertices will be: $\psi(4160)\to D_s^*D_s^*,\  DD_1,\ DD'_1\ $, to which correspond the 3-level interacting Lagrangian among a vector and two vectors 
\be
\label{l1}
\mathcal{L}_1=ig_{\psi VV}\ \Psi_{\mu\nu}\Big(D_{s}^{*\mu}\bar{D}_{s}^{*\nu} - D_{s}^{*\nu}\bar{D}_{s}^{*\mu} \Big),\ \ \Psi_{\mu\nu}=\partial_\mu \psi_\nu-\partial_\nu\psi_\mu\ ,
\ee
and among a vector and a pseudoscalar and an axial-vector or a pseudovector
\begin{align}
&\mathcal{L}_2=ig_{\psi PAV}\ \psi_\mu\Big(D\bar{D}_{AV}^\mu - D_{AV}^\mu\bar{D}\Big)\ ,\\
&\mathcal{L}_3=g_{\psi PPV}\ \psi_\mu\Big(D\bar{D}_{PV}^{\mu} + D_{PV}^{\mu}\bar{D}\Big)\ ,
\end{align}
where the physical states are
\begin{align}
&D_1=\cos\theta D_{AV}-i\sin\theta D_{PV}\ ,\\
&D_1'=\cos\theta D_{PV}-i\sin\theta D_{AV}.
\end{align}
For the second process we will have: $\psi(4415)\to D_sD_{s1},\  D^*D_1,\ D^*D'_1$, with the Lagrangians
\begin{align}
&\mathcal{L}_4=ig_{\psi PAV}\ \psi_\mu\Big(D_s\bar{D}_{sAV}^\mu - D_{sAV}^\mu\bar{D}_s\Big)\ ,\\
&\mathcal{L}_5=g_{\psi PPV}\ \psi_\mu\Big(D_s\bar{D}_{sPV}^\mu + D_{sPV}^\mu\bar{D}_s\Big)\ ,
\end{align}
where the physical states are given by
\begin{align}
&D_{s1}=\cos\theta D_{sAV}-i\sin\theta D_{sPV}\ ,\\
&D_{s1}'=\cos\theta D_{sPV}-i\sin\theta D_{sAV}\ ,
\end{align}
and between a vector, and a vector and an axial-vector or a pseudovector
\begin{align}
&\mathcal{L}_6=g_{\psi VAV}\ \Psi_{\mu\nu}\Big(D^{*\mu}\bar{D}_{AV}^{\nu} + D^{*\nu}\bar{D}_{AV}^{\mu} \Big)\ ,\\
&\mathcal{L}_7=ig_{\psi VPV}\ \Psi_{\mu\nu}\Big(D^{*\mu}\bar{D}_{PV}^{\nu} - D^{*\nu}\bar{D}_{PV}^{\mu} \Big)\ ,
\end{align}
where $\Psi$ is defined in Eq.~\eqref{l1}. After solving the amplitude corresponding to these interactions, one can evaluate the partial decay widths using
\be
\Gamma_j(s)=\frac{1}{8\pi}\frac{p_j(s)}{s}|\mathcal{M}_j|^2\ ,
\ee
where $|\mathcal{M}_j|^2$ and $p_j$ are the invariant amplitude squared and relativistic momentum for channel $j$, respectively. Now, we wish to include all the 3-level meson-meson 1-loops in a coupled-channel manner, according to the diagram in Fig.~\ref{fig2}.

The complex self-energy is given by
\be
\Sigma(s)=\sum_j^N \Big( \Omega_j(s) + i\sqrt{s}\Gamma_j(s)\Big),\ \ \Omega,\ \Gamma \in \Re,
\ee
where $N$ is the number of decay channels. The real part $\Omega$ can be computed using the dispersion relations, i.e.,
\be
\Omega_j(s)=\frac{1}{\pi}\int_{s_{th}}^\infty\frac{\sqrt{s'}\Gamma_j(s')}{s'-s}\ \ds s'.
\ee
And the propagator, for a single $\psi$, will be
\be
\label{prop}
\Delta(s)=\frac{1}{s-m_\psi^2+\Sigma(s)}.
\ee
The  unitarized spectral function, as a function of energy $E$, is proportional to the imaginary part of the propagator, where the denominator includes all channels
\be
\label{sf}
d_{\psi}(E) =-\frac{2E}{\pi}\mathrm{Im}\ \Delta(E)=\frac{2E^{2}}{\pi}\frac{\sum_j^N\Gamma_j(E)}{[E^{2}-m_{\psi}%
^{2}+\sum_j^N\Omega_j(E)]^2+[E\sum_j^N\Gamma_j(E)]^2}\text{ }.
\ee

The line-shape to each decay channel might be revealing, and thus motivate the experiment. As it may be seen in Table \ref{table}, the $\psi(4160)$ and $\psi(4415)$ lie only about 33 MeV and 13 MeV below the $D_s^*D_s^*$ and $D_sD_{s1}$ thresholds, respectively, yet actually, if we take into account their widths, each $\psi$ merge with the thresholds, therefore, the tail of each $\psi$ is expected to be clearly seen to those channels. The same idea appears in Refs.~\cite{prd76p074016} and \cite{epja36p189}, where a structure seen in channel $DD$ would be the tail of an hypothetical scalar charmonium at 3.7 GeV, or the case of the famous axial vector $X(3872)$, likely to be a loosely bound state below threshold, that leaves a fair tail in channel $DD^*$, see Refs.~\cite{epj71p1762} and \cite{epja36p189}.


\section{The case of the $\psi(3770)$}

\begin{figure}[t]
\centering
\resizebox{!}{200pt}{\includegraphics{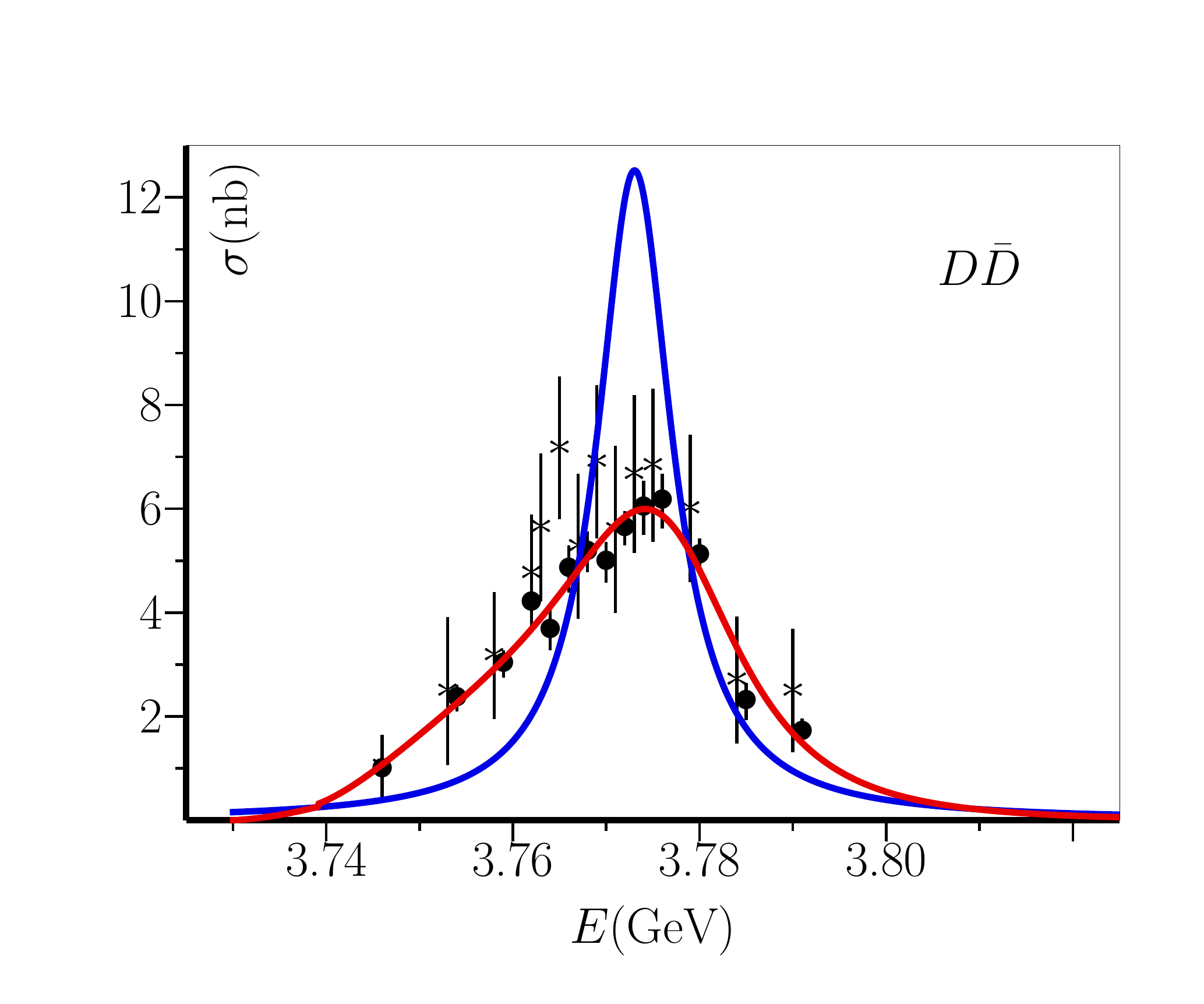}}
\mbox{}\\[-0mm]
\caption{\label{fig3}Data from BES, $e^+e^-\to D\bar{D}$, in Refs.~\cite{prl97p121801}-\cite{plb668p263}. Red line: fit to data in Ref.~\cite{paper}. Blue: theoretical Breit-Wigner without loop corrections.}
\end{figure}

In Refs.~\cite{1708.02041,paper} we studied the line-shape of the $\psi(3770)$ in channel $D\bar{D}$. We verified that the interference between this resonance and the $D^0\bar{D}^0$ and $D^+D^-$ thresholds was not only kinematic but also dynamical, through 1-loop contributions, in a similar manner to Fig.~\ref{fig2}. We computed the cross section $\sigma_{D\bar{D}}=\sigma_{D^0\bar{D}^0}+\sigma_{D^+D^-}$ and fitted to data in Refs.~\cite{prl97p121801,plb668p263} with four parameters. In Fig.~\ref{fig3} we show our result, that we compare with the cross section of a Breit-Wigner-like shape, computed with the same parameters, but without loop corrections. 
In this case, the interaction lagrangian was
\be
\mathcal{L}_{\psi(3770)\to D\bar{D}}=ig_{\psi D\bar{D}}\psi_\mu\Big(\d^\mu D\bar{D} - \d^\mu \bar{D}D \Big)
\ee
leading to a cross section given by
\be
\sigma_{e^{+}e^{-}\rightarrow \dd}=\frac{\pi}{2E}g_{\psi e^{+}e^{-}}%
^{2}d_{\psi}(E)=-g_{\psi e^{+}e^{-}}^{2}\mathrm{Im}\Delta(E)\text{
,}\label{cs}%
\ee
where the expressions for $d_{\psi}(E)$ and $\mathrm{Im}\Delta(E)$ are similar to Eqs.~\eqref{sf} and \eqref{prop} (see Ref.~\cite{paper} for all the details). Moreover, we examined the pole structure, and found two poles, at about $3777-i12$ MeV and $3741-i19$ MeV, the first one coming the ``seed'' and responsible for the main structure of the resonance, and the second one generated dynamically and partially responsible for the deformation of the signal. A similar behavior, within the same approach, can be found for the scalar kaonic system in Ref.~\cite{npb909p418}.

\section{Summary}

We analyze new vectorial enhancements measured by the BES Collaboration on the charmonium energy region, namely the $Y(4220)$, $Y(4260)$, $Y(4360)$, and $Y(4390)$, and we argue they might not be true resonances, since the continuum was not taken into account in the fits. An unitarized effective Lagrangian model is suggested, to study the interplay among the $\psi(4160)$ and channels $D_s^*D_s^*$, $DD_1$, and $DD_1'$, and among the $\psi(4415)$ and channels $D_sD_{s1}$, $D^*D_1$, and $D^*D_1'$, from which we expect to analyze line-shapes and poles. We present our recent result on the $\psi(3770)$, where the one-loop $D\bar{D}$ distorts the line-shape and originates a second dynamically generated pole. Such interesting results motivate us to extend the model to study other resonances. 

\section*{Acknowledgements}
We thank to Francesco Giacosa for very important discussions.
This work was supported by the \textit{Polish National Science Center} through the project OPUS no.~2015/17/B/ST2/01625.


\begin{thebibliography}{99}

\bibitem{prl118p092001}
M.~Ablikim {\it et al.} (BES Collaboration),
\href{https://doi.org/10.1103/PhysRevLett.118.092001}{\ Phys. Rev. Lett. {\bf 118}, 092001 (2017).}

\bibitem{prl118p092002}
M.~Ablikim {\it et al.} (BES Collaboration),
\href{https://doi.org/10.1103/PhysRevLett.118.092002}{\ Phys. Rev. Lett. {\bf 118}, 092002 (2017).}

\bibitem{pdg}
C. Patrignani {\it et al.} (Particle Data Group),
Chin. Phys. C 40, 100001 (2016).

\bibitem{prd79p111501}
E.~van Beveren and G.~Rupp,
\href{https://doi.org/10.1103/PhysRevD.79.111501}{\ Phys. Rev. D {\bf 79}, 111501(R) (2009).}

\bibitem{prl105p102001}
E.~van Beveren, G.~Rupp, and J.~Segovia,
\href{https://doi.org/10.1103/PhysRevLett.105.102001}{\ Phys. Rev. Lett. {\bf 105}, 102001 (2010).}

\bibitem{epja36p189}
D.~Gamermann and E.~Oset,
\href{https://doi.org/0.1140/epja/i2007-10580-5}{\ Eur. Phys. J.A {\bf 36}, 189 (2008)}.

\bibitem{1708.02041}
S.~Coito and F.~Giacosa,
\href{https://arxiv.org/abs/1708.02041}{arXiv: 1708.02041 [hep-ph]}.

\bibitem{paper}
S. Coito and F. Giacosa,
\href{}{arXiv:1712.00969 [hep-ph]}.

\bibitem{fpaper}
In preparation.

\bibitem{ppnp67p449}
G. Rupp, S. Coito, and E. van Beveren
\href{https://doi.org/10.1016/j.ppnp.2012.01.009}{Prog. Part. Nuc. Phys. {\bf 67}, 449 (2012)}.

\bibitem{prd77p011103}
G. Pakhlova {\it et al.} (Belle Collaboration)
\href{https://doi.org/10.1103/PhysRevD.77.011103}{\ Phys. Rev. D {\bf 77}, 011103(R) (2008).}

\bibitem{prl95p142001}
B.~Aubert {\it et al.} (BaBar Collaboration)
\href{https://doi.org/10.1103/PhysRevLett.95.142001}{\ Phys. Rev. Lett. {\bf 95}, 142001 (2005).}

\bibitem{prd76p074016}
D. Gamermann, E. Oset, D. Strottman, and M.J. Vicente Vacas,
\href{https://doi.org/10.1103/PhysRevD.76.074016}{\ Phys. Rev. D {\bf 76}, 074016 (2007).}

\bibitem{epj71p1762}
S. Coito, G. Rupp, E. van Beveren,
\href{https://doi.org/10.1140/epjc/s10052-011-1762-7}{\ Eur. Phys. J.C {\bf 71}, 1762 (2011)}.

\bibitem {prl97p121801}
M. Ablikim, \textit{et al.} (BES Collaboration),
\href{https://journals.aps.org/prl/pdf/10.1103/PhysRevLett.97.121801}{Phys. Rev. Lett. \textbf{97}%
, 121801 (2006).}

\bibitem {plb668p263}
M. Ablikim, \textit{et al.} (BES Collaboration),
\href{https://doi.org/10.1016/j.physletb.2008.08.067}{Phys. Lett. B \textbf{668}, 263 (2008)}.



\bibitem{npb909p418}
T.~Wolkanowski, M.~So\l tysiak, and F.~Giacosa,
\href{http://doi.org/10.1016/j.nuclphysb.2016.05.025}{Nuc. Phys. B
\textbf{909}, 418 (2016).}

\end{thebibliography}
\end{document}